\documentclass{acmart}

\acmJournal{TOMS}
\acmVolume{49}
\acmNumber{2}
\acmArticle{1}
\acmYear{2023}
\acmMonth{6}

\setcopyright{acmcopyright}
\copyrightyear{2022}

\acmDOI{0000001.0000001}

\usepackage{amsmath}
\usepackage{amsfonts}
\usepackage{hyperref}
\usepackage{todonotes}

\usepackage{listings}
\usepackage{tcolorbox}
\tcbuselibrary{skins,breakable}
\newtcolorbox{codebox}[2][]{
    colback=white,
    colframe=blue,
    coltitle=black,
    breakable = true,
    boxsep = -2.5mm,
    fonttitle=\small\ttfamily,enhanced,
    attach boxed title to top right={xshift = -0.5cm, yshift = -5.5mm},
    boxed title style={colback=white, colframe=blue},
    title=#2,#1
}

\newcommand{\dealii}{\texttt{deal.II}}

\newcommand{\petsc}{\texttt{PETSc}}
\newcommand{\trilinos}{\texttt{Trilinos}}

\begin{document}
% Title portion
\title{Experience converting a large mathematical software package written in C++ to C++20 modules}

% \subtitle{This is a subtitle}
% \subtitlenote{Subtitle note}

\author{Wolfgang Bangerth}
\orcid{0000-0003-2311-9402}
\affiliation{%
  \department{Department of Mathematics and Department of Geosciences}
  \institution{Colorado State University}
  \city{Fort Collins}
  \state{CO}
  \country{USA}}
\email{bangerth@colostate.edu}

\renewcommand\shortauthors{W. Bangerth}

\begin{abstract}
Mathematical software has traditionally been built in the form of ``packages'' that build on each other. A substantial fraction of these packages is written in C++ and so describe their interface via header files that downstream packages can then ``\texttt{\#include}'' -- an approach that C++ has inherited C, but that is considered clunky, unreliable, and slow. C++20 has introduced a ``module'' system in which packages explicitly export entities that compilers then store in machine-readable form and that downstream users can ``import''.

Herein, I explore approaches towards conversion of large mathematical software packages to modules, using the widely-used finite element library deal.II with 800,000 lines of code as an example. I describe how one can create both header- and module-based interfaces from the same code base, discuss the challenges one encounters, and provide practical experience with regard to technical and human metrics. With a non-trivial, but not prohibitive effort, the conversion to modules is possible, resulting in shorter compile times for the converted library; for downstream projects, compile times show no clear trend. I end with thoughts about long-term strategies for converting the entire ecosystem of mathematical software over the coming years or decades.
\end{abstract}






\maketitle

\section{Introduction}

A sizable fraction of the world's mathematical software is written in C++. In particular, this is true for the large foundational software libraries that underlie many scientific applications; for example, among many others, the following widely used packages are all written in C++: Eigen \cite{eigen-web-page} and Trilinos \cite{heroux2005trilinos,trilinos-web-page} for dense and sparse linear algebra; finite element packages such as dolphin \cite{alnaes2015}, MFEM \cite{anderson2021}, DUNE \cite{bastian2021}, libMesh \cite{kirk2006}, and deal.II \cite{dealII2020design,dealii96}; and the IPOPT \cite{ipopt} optimization package. All of these packages have roots that date back at least a decade and consequently utilize C++'s approach -- inherited from C -- of using header files to provide declarations (and in some cases implementations) to downstream users of classes and functions, and source files for their implementation. Based on these packages, thousands of applications, also written in C++, interface with the underlying packages via these header files.

C++ header files have a well-earned reputation for being clunky and slow. This is due to the fact that \textit{everything} that is listed in a header file is copied verbatim to the scope where the header file is included, and that the cumulative inclusion of headers can amount to hundreds of thousands -- and in some cases, millions -- of lines of code a compiler has to read, parse, and understand before a file's content can be compiled. As a consequence, nearly every language developed over the past several decades has a proper module system, from Modula in 1975 to Fortran 90 (1990), Python (1991), or Java (1995). C++ finally gained support for modules that do not rely on textual inclusion of other files with the C++20 standard, but it has taken until late 2024 for compilers and in particular the tool chain (e.g., the CMake configuration management system and the Ninja build tool) to make this feature functionally available in a platform independent way. 

It is clear that much of the mathematical software written in C++ -- both foundational libraries as well as the applications on top of these libraries -- will eventually want to switch to the use of modules over header files, perhaps following a phase where libraries offer access to their functionality through both. This is because modules offer a saner approach towards describing what a package wants to export, and because it has the potential to substantially accelerate the compilation process of C++ code. While it is not difficult to find blog posts, StackExchange questions, and other pages on the internet about the basic approach to C++20 modules, along with small toy examples, there is very little practical experience on how large-scale software packages such as those used in scientific computing can be converted to the use of C++20 modules. Indeed, the requirements for converting code bases with hundreds of thousands or millions of lines of code are very different than writing small projects from scratch: First, there are often thousands of users whose compatibility requirements have to be considered; second, it is impractical to re-write even small fractions of these code bases to accommodate a new paradigm. Converting such codes has to take these realities into account, and one has to come up with strategies that result in the desired outcome while keeping backward compatibility and feasibility in mind.

This paper presents my experience in implementing C++20 modules for the large-scale finite element library \dealii{}. Like many other large-scale software packages, \dealii{} dates back to the 1990s and today consists of some 800,000 lines of C++ code, along with about as much again in the test suite. Its code base has substantially evolved over the decades and now relies on and uses modern C++ features throughout, requiring a C++17 compiler since version 9.6, released in 2024. It is widely used in computational science, with thousands of users around the world; a list of more than 2800 publications that use \dealii{} can be found at \url{https://dealii.org/publications.html}. Given the size and scope of the library, it is clear that a whole-sale rewrite -- or even just substantial modifications to each of its 652 header and 399 implementation files -- is not feasible. Yet, based on implementing a hobby project of mine (the SampleFlow library) as part of teaching myself C++20 modules during the Thanksgiving break in 2024, it seemed conceivable to me that with appropriate scripting and building on today's toolchains, \dealii{} could be converted to use modules. Furthermore, for SampleFlow, I observed that compiling the test suite using modules took only about 50\% of the time compared to using traditional header files -- a speed up worth pursuing given that large C++ projects often require on the order of several CPU hours to compile and test; for example, \dealii{} takes about an hour of CPU time to compile, and around 49 CPU hours to run the full test suite. As a consequence of this initial experiment, I set out to see what it would take to provide a module-based interface for \dealii{}, based on the following principles:
\begin{itemize}
    \item The software must continue to provide a header-based interface to preserve existing users' ability to use the library, and these headers must continue to be compatible with whatever C++ standard was previously used even if it predates C++20.
    \item The software should provide a module-based interface that users should be able to use with minimal effort.
    \item The conversion cannot require a significant manual rewrite of the existing implementation. In practice, that means that much of the work must be done by scripts that are applied to many or all header and source files, resulting in files that can serve as compiler input to produce a C++20 module.
    \item The conversion should work within the confines of widely used toolchains. In particular, in the case of \dealii{}, this means that it must rely on the CMake configuration management system, and use one of its widely used build system backends; at the time of writing, only Ninja and Microsoft Visual Studio are able to build modules based on CMake output.
    \item The implementation of any new code should be relatively confined to a few files so that it can reasonably be reviewed by other maintainers. Keeping the changes localized also allows readers of this article to see how the implementation works, as well as copy-paste the code into their own work; my implementation is licensed under a liberal open source license to make this possible.
    \item The implementation should, hopefully, provide benefits in terms of code modularity, compile time, or along other metrics.
\end{itemize}
This paper reports on my experience with this project, what design decisions I took along the way, how difficult or time consuming the work was, and what its outcomes were. The goal is to document these experiences so that others contemplating the conversion of their own projects can hopefully follow similar steps with the confidence that they can work out. The metrics I will use to assess the outcomes of my experiment, and which I will report on below in detail, are: (i) Can packages such as \dealii{} be converted to a module-based system? (ii) If so, how much work did that take, and how technically involved was this work? (iii) How does the switch to modules affect compile time of \dealii{} and of downstream packages.

In order not to build too much suspense, let me also briefly preview the results of this experiment: (i) It was possible, though with an effort of perhaps 1.5 months of full-time work, to convert the code base to a module-based system; (ii) for the library itself as well as for some applications, compile times are substantially shorter whereas for a few applications, compile times are longer; (iii) it is not entirely clear at this point how libraries built into modules should communicate module interfaces to downstream applications through configuration systems such as CMake. All of these points will be substantiated in detail in the following pages.

\paragraph*{How does this relate to mathematical software} It is true that the experiment described herein has a component that is simply software engineering related to using new language features. At the same time, mathematical (and scientific software in general) differs from the broader software world in ways that have direct bearing on this conversion. Specifically, most scientific applications -- including mathematical software libraries such as those mentioned in the first paragraph -- are written by relatively small groups that build on a deep hierarchy of other libraries that are out of their control. In contrast, many companies today use ``monorepos'' in which the vast majority of code is available for change in a central place, under the control of the company's programmers \cite{brousse19}.%
\footnote{Of course, different parts of a monorepo are ``owned'' and maintained by different teams (though there are exceptions, see \cite{Tang_2025}), but the model still implies a far greater ability for downstream components to affect upstream code if that is necessary to achieve a goal; similarly, upstream components wanting to enact incompatible changes can do so because they have access to all or most downstream users \cite{Potvin2016}. In contrast, while a \dealii{} developer can submit patches to, say, \petsc{}, there is no leverage to actually get this patch approved. Moreover, we can not rely on this patch being present in any given setting: It will typically require the \petsc{} project to publish a release, and for this release then to be packaged by distributions. The \dealii{} code base contains many places where we have to support interfaces of external dependencies that were deprecated or removed years ago, but that are still found on our users' systems.}
To illustrate this difference, consider the example of the Advanced Solver for Planetary Evolution, Convection, and Tectonics (ASPECT, \cite{KHB12,HDGB17}). This code of currently 197,000 lines of code builds on \dealii{} (797,000 lines of code) for finite element functionality, which in turn builds on some 30 other libraries (see \cite{dealii96} for the full list) that include Trilinos (3,085,000 lines) and PETSc (764,000 lines) for parallel linear algebra, and BOOST (3,338,000 lines) for general tools. All of these, in turn, rely on yet other packages. In other words, ASPECT builds on external libraries that, combined, likely have about two orders of magnitude more code than the application itself. This deep and broad web of dependencies is uncommon in commercial software \cite{p2409r0}, but entirely typical of mathematical software. In practice, it means that for mathematical software packages, \textit{dealing with external dependencies is an important and unavoidable aspect}; the example also shows that it is not a \textit{small} aspect -- in fact, as will become clear in Section~\ref{sec:practical-aspects-external-headers}, dealing with external dependencies turned out to be a major component of the work on this project. As a consequence, I believe that this paper has direct bearing on how we develop \textit{mathematical} software, beyond the question of how one can use C++20 modules in software in general, simply because mathematical software is \textit{different} from the much larger realm of most commercial software.

\paragraph*{Overview}
The remainder of the paper is structured as follows: Section~\ref{sec:module-outline} gives a brief overview of how C++20  modules are designed by the C++ standard. Section~\ref{sec:principles} then outlines some of the general principles one likely needs to follow when converting large code bases to C++20 modules. Section~\ref{sec:steps} then discusses the concrete steps I had to take to provide a module for \dealii{}, including how one deals with header files, source files, external dependencies, configuration management, and how all of this is described in the CMake configuration and build system. Section~\ref{sec:experiences} then discusses how the conversion using these concrete steps actually proceeded in practice, and how the resulting software works in reality. I conclude in Section~\ref{sec:conclusions} with a summary of my findings and a discussion of implications for the continued evolution of the mathematical software ecosystem.

\paragraph*{References for terms frequently used throughout this paper}
Many parts of the remainder of this paper will reference the 2020 revision of the C++ standard that introduced modules. Formally, this standard is called ISO/IEC 14882:2020. Rather than annotate each of these places with a reference, I will simply refer to it once here \cite{cpp20}. This standard is also the normative source for how modules should work; any deviation in the descriptions below is my mistake. For non-normative sources, I found the Clang++ overview of modules at \url{https://clang.llvm.org/docs/StandardCPlusPlusModules.html} useful, as well as the more technical specification on cppreference at \url{https://en.cppreference.com/w/cpp/language/modules}.
Throughout this paper, I will also frequently reference a number of other mathematical software libraries, specifically \petsc{} \cite{petsc-user-ref,petsc-web-page} and \trilinos{} \cite{heroux2005trilinos,trilinos-web-page}, two large and widely-used libraries for parallel linear algebra, linear and nonlinear solvers, and other components of scientific codes; and Boost \cite{Boost-web-page}, a library of widely-used general-purpose C++ classes and functions. I will omit providing references for these libraries in the following in the interest of brevity, but supply references for other libraries mentioned in only a specific context.

\section{A very brief overview of C++20 modules}
\label{sec:module-outline}

A slightly simplified perspective is that in traditional C++ (as in C), software packages describe their application programming interfaces (APIs) by placing \textit{declarations} of classes, functions, variables, and other entities into ``header files''. The implementation of functions (their \textit{definition}) is provided separately in a ``source file'' that compilers then compile into static or shared libraries.%
\footnote{Some packages are ``header-only'' in that the implementations of functions are \textit{also} in the header files, and there is no compiled library. One can also place ``inline'' function definitions into header files, and often has to do that with template functions.
In the following, we will simply consider these as special cases of the situation outlined in this section and not consider it as different.}
Applications building on such a package tell the compiler what classes, functions, and variables are available from the underlying package by including its header files via a \texttt{\#include} statement -- in essence a literal copy of the underlying package's header file into the application's source file. The textual inclusion of header files -- which can amount to hundreds of thousands of lines -- is a major factor for the slowness of compiling C++. It also makes it quite difficult for package authors to control in a fine-grained way which information they want to export or not; for example, types only used in \texttt{private} parts of a class still need to be declared in header files, even though one might not actually wish to make them usable by downstream applications.

Modules address both of these drawbacks. They are intended to make it possible for package authors to explicitly state what it is they want to \textit{export} from a module file, i.e., which declarations, inline functions, templates, and other entities form the public interface of the package. The compiler would then produce a machine-readable description of the exported information of a module, which an application file (or a different part of the package itself) can then import wholesale. Since importing a module does not require reading, parsing, and understanding the contents of human-readable text files, importing a module should be faster than including a header file.

Among the design decision of C++20 modules that will become relevant in the following is that an entire self-contained package constitutes a single module; unlike in some other languages, a module \textit{may} consist of the contents of a single file, but typically is built from many files. Moreover, each declared entity the compiler sees is either not part of a module, or is uniquely tied to a single module. Finally, modules cannot depend on each other cyclically, which implies in particular that classes that depend cyclically on each other need to be in the same module.

The simplest form of a module will typically look as follows:%
\footnote{Complete descriptions of the C++20 module syntax can be found in the references provided at the end of the introduction. Our goal here is to convey the \textit{ideas} that are necessary to understand the design choices made in converting mathematical software packages.}
\begin{codebox}[]{example\_module\_file.ccm}
\begin{lstlisting}
module;                // This file belongs to a module; the
                       // 'global module fragment' starts here

                       // Include or import declarations from elsewhere:
#include <system_header.h>
import other_package;

module LinearAlgebra;  // The module 'LinearAlgebra' starts here
export                 // Declaration of exported entities
{
  class Matrix {...};
  void matrix_matrix_product (const Matrix &left_factor,
                              const Matrix &right_factor,
                              Matrix       &result);
}

                       // Declarations and definitions that are not exported:
void helper_function (...);
void matrix_matrix_product (...) { ... implementation ...}
\end{lstlisting}
\end{codebox}
This file implements a module called \texttt{LinearAlgebra} that exports two entities, the \texttt{Matrix} class and a function called \texttt{matrix\_matrix\_product()}. The implementation of that function, along with a helper function, are also provided in the same file, but are not exported -- traditionally, they would of course have been in a separate source (rather than header) file. Alternatively, one could mark the \texttt{matrix\_matrix\_product()} function definition as \texttt{inline} and also export it, resulting in the same effect as having an inline function declared in a header file.

\subsection{Interface module units}
\label{sec:module-outline-interface-unit}

In practice, software libraries export hundreds, thousands, or even larger number of declarations. For example, \dealii{} has 652 header files that each declare between a single class with all its members, and dozens of functions.%
\footnote{The release mode shared library for \dealii{} exports more than 300,000 symbols. While many of these are for functions not intended to be part of the public interface, or compiler-generated symbols, the number is a useful upper bound for the number of exported declarations for \dealii{}.}
Together, they comprise about 520,000 lines of declarations, code, and documentation. It is not reasonably possible to put all of these into a single module file, and consequently C++20 provides a way to split a single module into several \textit{partitions}. Each partition is described by a single file; for historical reasons, the C++ standard calls files ``units'', and I will follow this nomenclature in the remainder of this manuscript.

Using the example above, we could have one file that describes the partition \texttt{partition\_Matrix} of the module \texttt{LinearAlgebra} and exports the matrix class:%
\footnote{By convention, some compilers expect \textit{interface} units to have a file suffix \texttt{.ccm}, whereas the \textit{implementation} units discussed in the next section use \texttt{.cc}. We will use this convention in the following.}
\begin{codebox}[]{module\_interface\_partition\_unit\_1.ccm}
\begin{lstlisting}
module;
[...]
export module LinearAlgebra :partition_Matrix;
export class Matrix {...};
\end{lstlisting}
\end{codebox}
Here, the \texttt{export} keyword in \texttt{export module LinearAlgebra :partition\_Matrix;} ensures that the contents of the module partition is exported into the public interface of the module; the \texttt{export} keyword before the class declaration ensures that the class is exported into the public interface of the module partition.

A separate file describes a separate module partition named \texttt{partition\_matrix\_matrix\_product} that imports the partition above and exports the matrix-matrix product function (as part of the same module):
\begin{codebox}[]{module\_implementation\_partition\_unit\_2.ccm}
\begin{lstlisting}
module;
[...]
export module LinearAlgebra :partition_matrix_matrix_product;
import :partition_Matrix;

export void matrix_matrix_product (const Matrix &left_factor,
                                   const Matrix &right_factor,
                                   Matrix       &result);
\end{lstlisting}
\end{codebox}
In the language of the C++20 standard, each of these two files are ``interface module units'' (because they export their respective module partitions) and are compiled separately. Since the second file imports information from a module partition provided by the first, the build system needs to enforce a dependency that the second file can only be compiled once the first has been compiled. The fact that module units can import partitions from other module units also makes clear that they can not depend on each other cyclically; this is an important restriction to which we will come back in Section~\ref{sec:experiences-human-side}.

This outline already gives an indication how one might map existing \textit{header} files of a project onto interface module partitions. We will discuss the actual approach in Section~\ref{sec:steps-header-files}.

\subsection{Implementation module units}
\label{sec:module-outline-implementation-unit}

Not every module partition needs to be importable -- because not everything in a code base needs to be known to the outside. In traditional C++, those parts would be in source files, rather than header files, and would simply be compiled into a library. But, because declarations (and consequently their implementations) are tied to a module, if a module exports the declaration of a function, its implementation also needs to be part of the module. C++20 supports this using  ``implementation module units'' that differ from interface module units only in that they describe a module partition without exporting it. For the example above, then, we would have a third file as follows:
\begin{codebox}[]{module\_implementation\_partition.cc}
\begin{lstlisting}
module;
[...]
module LinearAlgebra :implementation_matrix_matrix_product;  // Note: no 'export'
import :partition_Matrix;
import :partition_matrix_matrix_product;

// Function implementation, not exported:
void matrix_matrix_product (const Matrix &left_factor,
                            const Matrix &right_factor,
                            Matrix       &result)
{ ... the actual implementation ... }
\end{lstlisting}
\end{codebox}
The module partition still has a name, but because we have no intention of importing the partition, the name is immaterial as long as it is unique.%
\footnote{Strictly speaking, implementation units that are part of a module do not \textit{need} to define an implementation partition; implementation units that are not implementation partition units are called ``pure implementation units'' in \url{https://stackoverflow.com/questions/70818433/the-orthogonality-of-module-interface-implementation-units-and-partitions}. However, implementation partition units allow for being more specific about what they import, and so I chose to focus on these for the conversion described herein.}

As in the previous sub-section, this outline provides an indication how one might map existing \textit{source} files of a project onto implementation module partitions. My  actual approach is then discussed in Section~\ref{sec:steps-source-files}.

\subsection{The primary module interface unit}
\label{sec:module-outline-primary-unit}

Just like header files, interface module units are collections of declarations that can either be for internal or external use of a library. In the traditional approach, internal header files may simply not be copied into installation directories if the programmers were diligent enough to mark a subset of header files as installable by the build system. In the context of modules, one needs what the C++ standard calls a ``primary module interface unit'' that describes the public interface of a module by listing \textit{all} interface partitions. Using the example from above, we would then need a file that looks as follows:
\begin{codebox}[]{primary\_module\_interface\_unit.ccm}
\begin{lstlisting}
module;
export module LinearAlgebra;  // Note: 'export module' but no partition designation
export import :partition_Matrix;
export import :partition_matrix_matrix_product;
\end{lstlisting}
\end{codebox}
External projects that use `\texttt{import LinearAlgebra;}' to use the modularized package then get exactly those declarations that are exported by one of the partitions listed in this primary module interface unit.

\section{Principles of converting a large-scale software library to use C++20 modules}
\label{sec:principles}

A reality of working with large software projects is that for major transitions like the one we are considering here, even minor changes made to many files can take unacceptably large amounts of time. For example, putting a specific piece of code into the right place in each file (or adding necessary header files, as mentioned in Section~\ref{sec:experiences-human-side}) might take 20-30 seconds per file -- but doing this for all 1051 files of \dealii{} then will take approximately a full day of (extremely boring) work. Similarly, individually annotating every class or function we want to export from a module is not feasible for a project of this size, even if from a conceptual perspective it would perhaps be the right thing to do.

As a consequence of this kind of consideration, nearly everything we will discuss in the following must be done by scripts that are applied to all files of the same kind and hopefully achieve the transformation correctly for all of them. In practice -- see the discussion in Section~\ref{sec:experiences} -- this actually works reasonably well because header and source files map well onto interface and implementation module partitions. The scripts I wrote for this project are not extensive -- in total about 400 lines of Python, of which 2/3 are comments, copyright headers, and other non-code lines. (This line count does not include long lists of header files for each of \dealii{}'s external dependencies, see Section~\ref{sec:practical-aspects-external-headers}.) The main work in converting software to use modules then, clearly, is not in writing the conversion scripts, but in dealing with the exceptional cases, as well as with external dependencies. The following sections outline the broad steps that were necessary for the conversion from a technical perspective. I will discuss in Section~\ref{sec:experiences} how the script-based conversion works in practice.

Relying on scripts has an additional benefit: As stated in the first bullet point of the goals outlined in the introduction, an important consideration for all mathematical software libraries is that they must continue to provide header-based interfaces for a long time to come. One can't just edit the existing files to convert them to module files because then they stop being header-based files. Obviously, to facilitate continuing development, it is also not possible to just \textit{fork} the code base into a header-based and a module-based version because then the two branches will quickly diverge in functionality. Rather, using scripts that convert from a header- to module-based system makes it possible for projects to offer both systems while continuing to evolve. Because they will have to run every time one builds the library (or every time a file changes), these scripts must also be 100\% successful and 100\% reproducible: They cannot require an additional human step, and they cannot come up with different solutions every time they are called (as AI coding assistants often do, for example).

\section{Concrete steps in converting a large-scale software library to use C++20 modules}
\label{sec:steps}

Given the outline of how modules are described to a C++ compiler in Section~\ref{sec:module-outline} and the principles for converting software projects of Section~\ref{sec:principles}, let me then discuss how the conversion was implemented in practice in this section. In the following, I will first explain how header files are converted to interface module partitions (Section~\ref{sec:steps-header-files}) and the primary module interface unit (Section~\ref{sec:steps-primary-module-interface-unit}), and then how source files are converted to implementation module partitions (Section~\ref{sec:steps-source-files}). Section~\ref{sec:steps-config.h} covers the special case of the \texttt{config.h} file that many projects use to describe configuration details -- say, the version number of the project, which other projects they interface with, etc. Sections~\ref{sec:practical-aspects-external-headers} and \ref{sec:steps-alternative} discuss what turned out to be a major obstacle: Dealing with the external dependencies that are so important in mathematical software, as discussed in the introduction. Finally, I will discuss how all of this is described to the CMake build system (Section~\ref{sec:steps-CMake}).

\subsection{Converting header files}
\label{sec:steps-header-files}

With the exception of the \texttt{config.h} file that is used by many software projects to record details of the configuration of a package (and which I will discuss in Section~\ref{sec:steps-config.h} below), a typical header file of \dealii{} looks as follows (I will omit the preprocessor guard against multiple inclusion and the copyright notice for brevity):
\begin{codebox}[]{base/utilities.h}
\begin{lstlisting}
#include <deal.II/base/types.h>
#include <list>

DEAL_II_NAMESPACE_OPEN
[...]                  // the declarations of utilities.h
DEAL_II_NAMESPACE_CLOSE
\end{lstlisting}
\end{codebox}
Here, \texttt{DEAL\_II\_NAMESPACE\_OPEN} is a macro that expands to `\texttt{namespace dealii \{}', with \texttt{DEAL\_II\_NAMESPACE\_CLOSE} correspondingly expanding to `\texttt{\}}'. These macros were introduced around 2004 in response to a user complaining about a name collision between a \dealii{} function and a function from a different package; at the time, \dealii{} did not use its own namespace and the macro seemed like a good compromise between using a namespace for users who wanted it, and retaining backward compatibility for those who did not. Later, we made the namespace the default, but left the macro because it allowed us to generate doxygen documentation that does not repeat the namespace prefix in hundreds or thousands of places.

Given the description of modules in Section~\ref{sec:module-outline}, we will need to translate the file above in some automated way into an interface module unit that looks as follows:
\begin{codebox}[]{base/utilities.ccm}
\begin{lstlisting}
module;
[...]
#include <list>
[...]
export module LinearAlgebra :interface_partition_base_utilities;
import :interface_partition_base_types;
export {
  DEAL_II_NAMESPACE_OPEN
  [...]                  // the declarations exported by this interface module partition
  DEAL_II_NAMESPACE_CLOSE
}
\end{lstlisting}
\end{codebox}
Achieving this transformation is not difficult with a small Python script that reads through the source file and performs the following actions:
\begin{itemize}
    \item Output the `\texttt{module;}' statement.
    \item For each line of the input file down to the occurrence of the \texttt{DEAL\_II\_NAMESPACE\_OPEN} macro, output the line as-is unless it \texttt{\#include}s a \dealii{} file; in the latter case, do not copy the line into the output, but instead add the name of the file to a list of included intra-project headers. (All remaining \texttt{\#include} statements then refer to extra-project header files and are copied verbatim -- but see Section~\ref{sec:practical-aspects-external-headers}.)
    \item Output the `\texttt{export module <name> :<partition>;}' statement, where \texttt{<name>} is the module name for the current project, and \texttt{:<partition>} is the name of the interface partition that is derived in a systematic way from the name of the header file currently being processed. In the example above, the partition name starts with `\texttt{interface\_partition\_}', followed by the path to the header file with slashes replaced by underscores, and with the header suffix `\texttt{.h}' removed.
    \item For each of the previously collected intra-project header inclusions, output `\texttt{import :<partition>;}'. The name of the imported partition is derived in the same systematic way from the name of the included header file as in the previous bullet point.
    \item Output the text `\texttt{export \{}' to start the section of exported declarations.
    \item Output \texttt{DEAL\_II\_NAMESPACE\_OPEN}.
    \item For each line of the input file down to and including the occurrence of the \texttt{DEAL\_II\_NAMESPACE\_CLOSE} macro, output the line as-is.
    \item Output the text `\texttt{\}}' to end the section of exported declarations.
    \item For each remaining line of the input file, output the line as-is.
\end{itemize}
The first version of the Python script that does that had 38 lines of actual code, plus comments and documentation.%
\footnote{The script was later refactored in view of the changes necessary for dealing with external projects, see Section~\ref{sec:practical-aspects-external-headers}. It is still quite small, but imports another Python script that contains long lists of header files corresponding to external projects \dealii{} interacts with.}
In the vast majority of the existing \dealii{} header files, this relatively straightforward transformation already yields a valid module interface partition unit. (Of course, in practice, even a small fraction of the 699 header files that trip up the script can cause a substantial amount of work -- see the discussion in Section~\ref{sec:experiences}.) It is clear that the fact that nearly every \dealii{} header file uses the \texttt{DEAL\_II\_NAMESPACE\_OPEN/CLOSE} macros is a lucky break that allows the script to inject text in just the right places of the header file. At the same time, many projects likely have similar structure that allows scripts to key certain actions off of other pieces of information that is present in all or most files.

The approach outlined above works. The key issue one runs into is a semantic difference between header files and interface module units: The former re-export transitively whichever other header files they include. In other words, given that the header file \texttt{deal.II/base/utilities.h} has an \texttt{\#include <deal.II/base/types.h>} statement, a file \texttt{example.cc} that has a \texttt{\#include <deal.II/base/utilities.h>} statement also has access to the declarations in the header file \texttt{<deal.II/base/types.h>}. This is not the case in the example above: importers of the \texttt{:interface\_partition\_base\_utilities} partition do not automatically also gain access to the declarations in partition \texttt{:interface\_partition\_base\_types}.

This leads to a very long list of compilation errors. At least in an abstract sense, the correct approach is to add \texttt{\#include <deal.II/base/types.h>} to \texttt{example.cc} (which will then be turned into \texttt{import :interface\_partition\_base\_types;} by scripts) if \texttt{example.cc} indeed uses declarations from \texttt{deal.II/base/types.h}. This is conceptually correct because every file \textit{should} explicitly include whatever declarations it uses, rather than relying on transitive inclusion of chains of header files.

In practice, however, being conceptually correct can require adding hundreds or thousands of these missing \texttt{\#include} statements to the existing files (before conversion to module units), each of which may take perhaps 20 seconds to find the file corresponding to a compiler error, scrolling to the right location to add the \texttt{\#include}, and adding what was missing. Because compiler messages are typically spread over several lines, and reference the place where an undeclared function or class is used, rather than the place where the \texttt{\#include} is missing, I could also not find a way to script this step.%
\footnote{A reviewer kindly pointed me to the ``include-what-you-use'' tool \cite{include-what-you-use}, a compiler-based tool that ensures that if a file uses a symbol \texttt{X}, then it must also explicitly \texttt{\#include} the header file in which \texttt{X} is declared. While this tool is not accurate enough to completely automate this work, it nonetheless would likely have been able to avoid a lot of the work mentioned in the main text.}
After spending a rather boring half-day adding 247 include statements to 116 \dealii{} files, I came to the conclusion that being conceptually correct is not achievable in the short run. However, the C++ standard provides for a way to re-export imported module partitions. Instead of the above translation, we can write it as follows:
\begin{codebox}[]{base/utilities.ccm}
\begin{lstlisting}
module;
[...]
#include <list>
[...]
export module LinearAlgebra :interface_partition_base_utilities;
export import :interface_partition_base_types;                  // note the 'export' keyword
export {
  ...as before...
\end{lstlisting}
\end{codebox}
This form ensures that the \texttt{:interface\_partition\_base\_utilities} partition not only imports, but also re-exports the \texttt{:interface\_partition\_base\_types} partition in the same way as the \texttt{deal.II/base/utilities.h} header file re-exports the contents of the \texttt{deal.II/base/types.h} header file it includes. This approach allowed me to make forward progress on the overall project, though I later reverted to the former form and gradually added the remaining header inclusions where necessary.

As a final remark, for arcane reasons having to do with the two-phase nature of pre-processing and compiling C++ code, the authors of the text that describes C++20 modules have chosen to require that module statements exporting and importing partitions cannot be in parts of a file that are guarded by \texttt{\#ifdef ... \#endif} preprocessor directives. In practice, this means that the conversion script also has to get rid of the usual header include guards at the top and bottom of header files. Moreover, many \dealii{} files wrap external packages, but everything in these files is enclosed in statements such as \texttt{\#ifdef DEAL\_II\_WITH\_PETSC ... \#endif}. In these cases, it is important that the \texttt{DEAL\_II\_NAMESPACE\_OPEN/CLOSE} lines are \textit{outside} the guarded block so that after expansion by the script, the \texttt{export} and \texttt{import} statements are outside the guarded block as well.

\subsection{Creating the primary module interface unit}
\label{sec:steps-primary-module-interface-unit}

Each of the interface module units created from header files using the process of the previous section exports an interface partition. From these, the primary module interface unit (see Section~\ref{sec:module-outline-primary-unit}) creates the public interface of the module that one can then import via a `\texttt{import dealii;}' statement (or using whatever else the project's module name should be).

In principle, one could choose to convert internal header files not into interface partition units, but into implementation partitions (which, like all partitions, are importable by other units), thereby excluding them from the public interface of the module. At least in \dealii{}, we do not track which header files are public and which are internal -- they all get installed. In the current context where everything needs to be scriptable to the extent possible, it is also not feasible to go through all 652 header files and identify which provide public and which provide internal declarations.

As a consequence, all header files are converted into interface units, and the script that creates the primary module interface unit simply takes a list of all created interface partition units, parses them to find the `\texttt{export module dealii :<partition\_name>;}' line, and then outputs a file that looks as follows: 
\begin{codebox}[]{dealii.ccm}
\begin{lstlisting}
module;
export module dealii;
export import :interface_partition_algorithms_any_data;
export import :interface_partition_algorithms_general_data_storage;
export import :interface_partition_algorithms_named_selection;
[... several hundred more lines of that form ...]
\end{lstlisting}
\end{codebox}

The script that does this consists of only 14 lines of Python, plus documentation.

\subsection{Converting source files}
\label{sec:steps-source-files}

We can apply in essence the same scheme used for header files in Section~\ref{sec:steps-header-files} also  to the conversion of source files. In \dealii{}, source files generally look as follows:
\begin{codebox}[]{source/base/utilities.cc}
\begin{lstlisting}
#include <deal.II/base/utilities.h>
#include <list>

DEAL_II_NAMESPACE_OPEN
[...]                  // implementation of the classes and functions of utilities.cc
DEAL_II_NAMESPACE_CLOSE
\end{lstlisting}
\end{codebox}
We want to transform this file to a module implementation partition unit that looks as follows:
\begin{codebox}[]{source/base/utilities.cc}
\begin{lstlisting}
module;
[...]
#include <list>
[...]
module LinearAlgebra :implementation_partition_base_utilities;
import :interface_partition_base_utilities;
DEAL_II_NAMESPACE_OPEN
[...]                  // implementation of the classes and functions of utilities.cc
DEAL_II_NAMESPACE_CLOSE
\end{lstlisting}
\end{codebox}
The steps necessary to achieve this transformation are fundamentally the same as for header files (Section~\ref{sec:steps-header-files}), with the exception that (i) the module partition itself is not exported, (ii) imports or other intra-project module partitions are not re-exported, (iii) the material delimited by \texttt{DEAL\_II\_NAMESPACE\_OPEN/CLOSE} should not be enclosed in `\texttt{export \{ ... \}}'. As before, this is not difficult to achieve with a minor variation of the same, rather short Python script.

\subsection{Practical aspects: Dealing with configuration and other macros}
\label{sec:steps-config.h}

Many projects, \dealii{} included, record information determined during configuration in a file that is traditionally called \texttt{config.h}. Examples of what this file contains are the version number of the package, for each potential optional external dependence whether or not the external dependence was detected, information about compilers (e.g., whether the compiler has bugs that one needs to work around in the code base) and the system (e.g., whether the processor supports vectorization instructions), etc. Historically, all of this information is stored in preprocessor variables that either have values, or that are defined/undefined.

The key issue with this approach is that modules do not export preprocessor variables or macros: Modules are created based on what the \textit{compiler} sees \textit{after} the preprocessor is run. As a consequence, \texttt{config.h} cannot be converted into an interface partition unit if we want to use the preprocessor variables and macros used there in other files (either within the project, or in downstream projects); rather, it must remain a traditional header file that each interface and implementation partition unit must continue to \texttt{\#include} right after the `\texttt{module;}' statement at the top. In other words, the descriptions in Sections~\ref{sec:steps-header-files} and \ref{sec:steps-source-files} was not entirely correct: In reality, the conversion scripts mentioned there place `\texttt{\#include <deal.II/base/config.h>}' in the line following `\texttt{module;}'.

Conceptually, this is not difficult. In practice, however, this leads to some downstream work. The fact that preprocessor variables are not exported by modules implies that \textit{all} header files that define preprocessor variables and macros must be treated differently -- or, better, that \texttt{config.h} should be the \textit{only} file to define preprocessor symbols. In \dealii{}, however, several other header files did so. For example, \texttt{deal.II/base/exceptions.h} defined \dealii{}'s \texttt{Assert} and related macros that are used throughout the code basis. The rest of the code obtained these macros via `\texttt{\#include <deal.II/base/exceptions.h>}' -- but this will no longer work once \texttt{\#include} is replaced by `\texttt{import :interface\_partition\_base\_exceptions;}'. Instead, all definitions of preprocessor variables and macros have to either be centralized into \texttt{config.h}, or the header files in which they are declared have to be treated like \texttt{config.h} in that they cannot be converted to module interface partitions. In practice, I have separated all preprocessor definitions into a handful of files that are combined into a single \texttt{deal.II/macros.h} file that \textit{exclusively} consists of preprocessor \texttt{\#define}s and nothing else, and that can be \texttt{\#include}d by both \dealii{}'s own module partitions and by user code that builds on the \dealii{} module.

Given that preprocessor variables and macros do not play nice with modules, it may be worthwhile to use the opportunity to consider whether C++20 allows us to move away from using the preprocessor altogether. In many cases this is possible by using `\texttt{constexpr}' variables and functions that, like macros, evaluate at compile time. In practice, however, this is not enough: Preprocessor defines such as \texttt{DEAL\_II\_WITH\_TRILINOS} (defined if \dealii{}'s CMake files have detected an installation of Trilinos that we can interface with) are not only used within block scope (i.e., within functions) where `\texttt{constexpr}' variables and functions can be used, but they are also used to comment out or in whole parts of files that should or should not be visible to the compiler -- i.e., the preprocessor variable is used within namespace scope where `\texttt{constexpr}' variables and function cannot be used. As a consequence, we cannot yet move away from the preprocessor for many configuration management tasks with the current facilities of the C++ standard.

\subsection{Practical aspects: Header files of external projects and modules}
\label{sec:practical-aspects-external-headers}

Header files create problems beyond the issue with \texttt{config.h} discussed in the previous section. As discussed in the introduction, nearly every mathematical software package builds on something else -- oftentimes on deep stacks of \textit{other} packages -- and the interaction with these external dependencies typically happens through the header files of these external packages. In practice, this has turned out to not work well, and the following two subsections explain what the issue is, and how to address it.

\subsubsection{The problem: Compilers duplicate information}
\label{sec:practical-aspects-external-headers-problem}

At least in principle, C++ defines modules in a way so that they can interact well with external projects that have either not yet been modularized, or are written in C, using traditional \texttt{\#include} statements: It is entirely acceptable to include other files in the global module fragment (the space between the leading `\texttt{module;}' statement and the start of a module or module partition declaration). Specifically, this of course also means that including C++'s own header files in the global module fragment is acceptable and -- clearly -- necessary because module units like nearly every other file typically depend on the standard C++ library. 

In practice, however, this does not work well with current compiler technology. To understand why, let us consider the following simple example using traditional header files:
\begin{codebox}[]{a.h}
\begin{lstlisting}
#ifndef header_guard_a_h   // ensure this file is only included once
#define header_guard_a_h

#include <string>
inline std::string get_name(...) {...}

#endif    // ifndef header_guard_a_h
\end{lstlisting}
\end{codebox}
\begin{codebox}[]{b.h}
\begin{lstlisting}
#ifndef header_guard_b_h   // ensure this file is only included once
#define header_guard_b_h

#include <a.h>             // b.h includes a.h
#include <string>
inline std::string get_address(...) {...}

#endif    // ifndef header_guard_b_h
\end{lstlisting}
\end{codebox}
\begin{codebox}[]{x.cc}
\begin{lstlisting}
#include <a.h>
#include <b.h>
#include <iostream>

int main () {
  std::cout << get_name() << get_address() << std::endl;
}
\end{lstlisting}
\end{codebox}

Here, the ``header guards'' ensure that even though \texttt{a.h} is included twice from \texttt{x.cc} (once directly, and once via \texttt{b.h}), the preprocessor ultimately places only one copy of the file's contents into the compiler's input. Likewise, the compiler-provided header file \texttt{<string>} will use a similar scheme. At the end of the day, when \texttt{x.cc} is compiled, the compiler sees only one copy of every header file that is included directly or indirectly. Moreover, upon compilation, the resulting object file will only contain the code generated by compiling functions, plus references to those functions declared in header files that are actually used in the compiled code -- in other words, a small fraction of what is likely declared in \texttt{<string>} and whatever that file itself \texttt{\#include}s.

With modules, the situation is more complicated, however. After conversion to module partitions, the example above will look as follows:
\begin{codebox}[]{a.ccm}
\begin{lstlisting}
module;
#include <string>
export module myproject :interface_partition_a;
export inline std::string get_name(...) {...}
\end{lstlisting}
\end{codebox}
\begin{codebox}[]{b.ccm}
\begin{lstlisting}
module;
#include <string>
export module myproject :interface_partition_b;
import :interface_partition_a;
export inline std::string get_address(...) {...}
\end{lstlisting}
\end{codebox}
\begin{codebox}[]{x.cc}
\begin{lstlisting}
module;
#include <iostream>
module myproject;
import :interface_partition_a;
import :interface_partition_b;
int main () {
  std::cout << get_name() << get_address() << std::endl;
}
\end{lstlisting}
\end{codebox}

The issue here is that (i) we are compiling all three files separately, and (ii) when current compilers compile a file such as \texttt{a.ccm}, they essentially store the \textit{entire} content of the translation unit (i.e., the state after preprocessing, which contains included header files) in the output file. In other words, the resulting compiled module partition contains the contents of \texttt{<string>} and everything the latter file transitively includes.%
\footnote{This is necessary because the module partition may export templates that can be instantiated in other translation units (such as \texttt{x.cc}), requiring knowledge about the declarations and definitions that were available \textit{at the time the template was declared in the \texttt{a.ccm}} file -- rather than just the declarations that \texttt{a.ccm} itself declares and exports.} 
Of course, the compilation of \texttt{b.ccm} \textit{also} includes a copy of \texttt{<string>} and everything the latter file transitively includes. When we come around to compiling \texttt{x.cc}, reading the compiled module partitions that resulted from \texttt{a.ccm} and \texttt{b.ccm} then means that the compiler reads two copies of decent portions of the C++ standard library, and will keep these in memory, along with a representation of the \texttt{<iostream>} file and its dependencies included from \texttt{x.cc} itself.%
\footnote{One could imagine compiler designs in which the representations of header files stored in compiled interface module units are unified and only one copy is kept. Current compilers do not do this, however, principally because the representations of header files stored in compiled module units need not be the same. This is because one can define preprocessor variables before writing \texttt{\#include <string>} that affect the material being included. This may sound obscure but is, for example way many compilers recommend to switch on or off assertions in the standard library. For example, whether or not the preprocessor variable \texttt{\_GLIBCXX\_DEBUG} is defined determines whether the GCC compiler sees the debug or release versions of many standard library functions defined in C++ standard library header files.}

In practice, source files may depend on dozens or hundreds of other header files, all of which include C++ standard header files, requiring large amounts of memory and compute time to read the same information in over and over again. In the best case, this is only wasteful; in the worst case, it exceeds the machine's memory and the programmer's patience. To give a concrete example, the \dealii{} source file \texttt{source/base/utilities.cc} directly or indirectly depends on only 25 other \dealii{} header files (or interface units, after conversion), of course plus parts of the C++ standard library. Compiling the original, header-based file requires 3.9 CPU seconds and 426 MB of memory using the clang-20 compiler; compiling the version converted to modules requires 34 seconds and 4,764 MB -- in other words, about ten times longer and requiring ten times more memory. In total, 30 of \dealii{}'s converted 652 header files, and 70 of its 399 converted source files cannot be compiled because the compiler runs out of memory.%
\footnote{In practice, for clang-20, it isn't actually the memory that is exhausted, but the 31-bit space used to index certain symbols. The result is of course the same.}
Each of the files that can not be compiled imports between 115 and 190 other module partitions either directly or indirectly, and each imported partition itself stores a copy of a sizable part of the C++ standard library and perhaps other third party headers.

\begin{figure}
    \centering
    \includegraphics[width=0.95\linewidth]{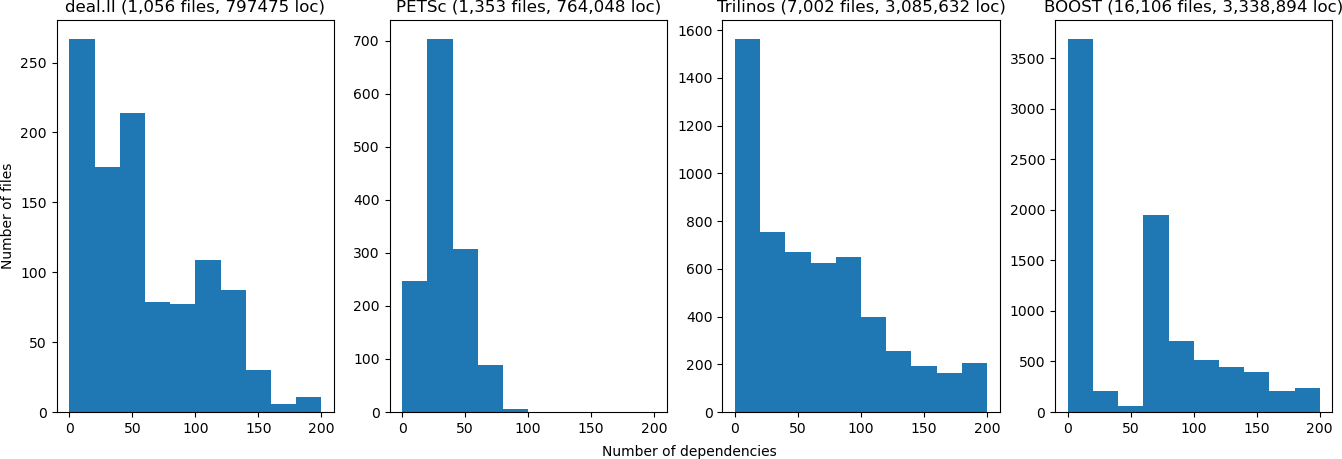}
    \caption{\it Histograms of how many other project-internal files each file in several mathematical software libraries depends on through \texttt{\#include} statements, either directly or indirectly. (Trilinos is best understood as a collection of sub-packages, each with its own wildly varying file naming and directory structure conventions. The histogram may contain some files that are not source but test or infrastructure files, though I have done my best to filter these out.)}
    \label{fig:dependencies}
\end{figure}

It is worth pointing out in this context that this is not a problem of \dealii{} alone. Fig.~\ref{fig:dependencies} shows histograms for the number of project-internal header files each file in \dealii{}, PETSc, Trilinos, and Boost respectively depend on. (Each of the project-internal header files then, in turn, may depend on an even larger number of system or C++ header files themselves.) From these data, it is clear that it is not uncommon for files to depend on 100 or more other headers. For example, Trilinos has a file that depends on 1309 other Trilinos files; Boost has a file that depends on 2491 other Boost files. One may infer that many other mathematical libraries written in C++ are written in a similar style. Under the scheme outlined so far, many of the source files of many of these projects would also not be compilable into C++20 modules.

\subsubsection{The solution: ``Wrapping'' external dependencies}
\label{sec:practical-aspects-external-headers-solution}

While the same header file included from different files can look very different based on what preprocessor symbols are set at the point of inclusion, importing the same module partition always results in the same outcome: If \texttt{b.ccm} has an \texttt{import :interface\_partition\_a;} statement, and \texttt{x.cc} has both an \texttt{import :interface\_partition\_a;} and an \texttt{import :interface\_partition\_b;} statement, then the representation of \texttt{:interface\_partition\_a} in the compiled versions of \texttt{b.ccm} and \texttt{x.cc} are the same because we are referencing a \textit{compiled} module partition that is fixed; specifically, it is fixed at the time of compiling the \textit{imported} partition, not at the time of compiling the \textit{importing} partition. As a consequence, the compiler will store only one copy of the imported partition when compiling \texttt{x.cc}, even though it is reached twice in the import graph.

This realization leads to a solution of the issue with external projects: namely, to ``wrap'' the declarations of the external project into a module partition that can be compiled once and imported multiple times. The mechanism for wrapping offered by the C++ standard is that one can write a \texttt{using} declaration within a module (partition) that, in contrast to ``traditional''  \texttt{using} declarations does not make a symbol available in a different namespace, but instead ties it to a module. For example, the functions and classes of \texttt{namespace std} are wrapped in the following way:%
\footnote{The C++23 standard provides that one can use \texttt{import std;}, which compilers would presumably implement in essence in the way shown here. However, as of the time of writing, compilers do not actually implement \texttt{import std;}, and CMake also only has rudimentary support for it. As a consequence, I did not attempt to use this feature; moreover, the approach shown here can also be used for other external dependencies, as will be discussed in the following.}
\begin{codebox}[]{std.ccm}
\begin{lstlisting}
module;

#include <algorithm>
#include <any>
[... list all other C++ headers ...]

export module dealii.external.std;

export
{
  namespace std
  {
    using std::int16_t;
    using std::int32_t;
    [... list more types ...]

    using std::abs;
    using std::acos;
    [... list more functions ...]

    using std::any;
    using std::array;
    [... list more classes ...]
  }
}
\end{lstlisting}    
\end{codebox}
The list of exported symbols here needs to encompass every C++ class, function, and variable that is used in the rest of the project. Given the size of \dealii{}, it is not surprising that this list has hundreds of entries (see Table~\ref{table:wrappers} below). In return, downstream module partitions can now simply write the following to gain access to all symbols in namespace \texttt{std} in one fell swoop:
\begin{codebox}[]{file\_that\_uses\_std.cc}
\begin{lstlisting}
  import dealii.external.std;
\end{lstlisting}
\end{codebox}
This process has to be repeated for all external dependencies. As it is for many other mathematical software projects, the list of dependencies of \dealii{} is long, and consequently many such wrappers have to be written; in practice this has turned out to be a tedious and substantial fraction of the overall work on this project. 

In the example above, I have put the wrapped symbols into a module \texttt{dealii.external.std} separate from that of \dealii{} itself. (Dots are part of the name of a module and carry no semantic meaning to the compiler.) One could have imagined instead using an interface partition that is part of the \texttt{dealii} module, but this has the downside that every importer of the \texttt{dealii} module also imports all wrapped symbols. This may be \textit{convenient} in some contexts, but is ultimately awkward because the list of wrapped symbols does not encompass \textit{every} symbol of the wrapped library but only those we have chosen to wrap (i.e., a superset of the symbols actually used within \dealii{} itself, but perhaps not including all symbols used in the downstream application). I considered it confusing if a user could use \textit{many} but not \textit{all} parts of \texttt{namespace std} this way, and consequently used a separate module for each wrapped library.

A principal complication for wrappers is that one can only export symbols that have ``external linkage''. While this happens to be true for (nearly) everything defined by the C++ standard, many other projects use symbols that fall into one of the following three categories that need to be wrapped differently:
\begin{itemize}
    \item \textit{Names with internal linkage:} C++ defines a number of ways in which names may have ``internal'' linkage within a translation unit. Examples are \texttt{static} functions, \texttt{const} and \texttt{constexpr} variables, or the names declared in an anonymous \texttt{enum}. Some libraries use these extensively. To name just one example, the MPI 4 standard \cite{mpi40} explicitly says that \texttt{MPI\_MAX\_ERROR\_STRING} is either a \texttt{const int} value or a member of an unnamed \texttt{enum}; in practice, many other MPI symbols like data type descriptors (e.g., \texttt{MPI\_DOUBLE}) are also defined as names with internal linkage by common MPI implementations. Names with internal linkage cannot be exported into a module with \texttt{using} statements. However, they can be wrapped in the following way (albeit not in their original namespace):
    \begin{codebox}[]{file\_that\_exports\_internal\_symbol.ccm}
    \begin{lstlisting}
namespace dealii {
  export const auto &MPI_MAX_ERROR_STRING = ::MPI_MAX_ERROR_STRING;
}
    \end{lstlisting}
    \end{codebox}
    Note that the new symbol has to be in \textit{our} project's namespace since we cannot re-declare the symbol in the global namespace with a different type or linkage. Like MPI, PETSc also extensively uses members of anonymous \texttt{enum}s that need to be wrapped like this.%
    \footnote{PETSc extensively makes use of the idiom \texttt{typedef enum \{ a, b, c \} E;} in which there is a named type, but the \texttt{enum} itself has no name and its members have internal linkage. Here, exporting the name \texttt{E} does not also export the \texttt{enum}'s members, and one has to export them individually as described in the main text.}

    \item \textit{Preprocessor variables:} Likewise, external libraries oftentimes define names in their header files as preprocessor symbols. These are not variables to the compiler, and so cannot can be referenced in \texttt{using} statements; however, they can be wrapped following a similar scheme as above (here using PETSc's \texttt{MATAIJ} -- denoting one way to represent a sparse matrix -- as an example):
\begin{codebox}[]{file\_that\_exports\_proprocessor\_variable.ccm}
    \begin{lstlisting}
namespace dealii {
  namespace internal {
    const auto petsc_mataij = MATAIJ;    // create a compiler variable under a different name
  }
#undef MATAIJ                            // make sure uses of the name are not replaced below
  export const auto &MATAIJ = internal::petsc_mataij;  // export a variable with the old name
}
    \end{lstlisting}
    \end{codebox}
    For this as for the previous case, the boilerplate code in this example can be abstracted using preprocessor macros.
    
    \item \textit{Preprocessor function-like macros:} The last category of symbol that cannot be exported is if external libraries define function-like preprocessor macros. For example, PETSc defines the following macro that can be used to check the return value of a call to a PETSc function:
\begin{codebox}[]{file\_that\_duplicates\_proprocessor\_macro.ccm}
    \begin{lstlisting}
#  define PetscCall(...)                                        \
    do                                                          \
      {                                                         \
        PetscErrorCode ierr_petsc_call_q_;                      \
        PetscStackUpdateLine;                                   \
        ierr_petsc_call_q_ = __VA_ARGS__;                       \
        if (PetscUnlikely(ierr_petsc_call_q_ != PETSC_SUCCESS)) \
          return PetscError(PETSC_COMM_SELF,                    \
                            __LINE__,                           \
                            PETSC_FUNCTION_NAME,                \
                            __FILE__,                           \
                            ierr_petsc_call_q_,                 \
                            PETSC_ERROR_REPEAT,                 \
                            " ");                               \
      }                                                         \
    while (0)
    \end{lstlisting}
    \end{codebox}
    Because it is a \textit{preprocessor} macro, it is invisible to the compiler proper. At the same time, it cannot be converted to a \textit{function} because its use relies on running, for example, \texttt{PetscStackUpdateLine} before the evaluation of the arguments provided through the ellipsis (``...''). Such macros can, therefore, not be exported at all. 

    This category of symbols thus presents a genuine conundrum: In order to use \texttt{PetscCall} (or any similar macro), we \textit{have} to \texttt{\#include <petscerror.h>} (the file in which the macro is defined) in every translation unit in which it is used. But \texttt{petscerror.h} also declares many other symbols, includes other PETSc headers and system header files, and consequently triggers the issues discussed in Section~\ref{sec:practical-aspects-external-headers-problem} above. The only solution I have found is therefore to duplicate the definition of \texttt{PetscCall} into a file that \textit{only} declares macros such as this, and nothing else. This is very clearly not a good solution, but the only one I could come up with.
\end{itemize}

As mentioned above, \dealii{} interfaces with \textit{many} other libraries -- a common feature of mathematical software. For this project, I therefore wrote wrappers to 21 other libraries, covering all of the major mandatory and optional dependencies \dealii{} has. Each of these wrappers tended to present its own set of complications. Generally, external projects fell into two categories: 
\begin{itemize}
    \item Modern C++ projects in which symbols are declared in a project-specific namespace were relatively straightforward to wrap. This is because they typically use a single namespace, making it easy to find, via \texttt{grep}, which symbols we are using in \dealii{}, and then to wrap them. Furthermore, these projects use relatively few classes one needs to export (which automatically also exports class members). Finally, they often use the preprocessor lightly. Examples for this category include Kokkos \cite{trott2022} and Taskflow \cite{huang2021taskflow}.
  \item External libraries written in C or in more historical C++ styles were more troublesome to wrap. This is because they have a large number of global names that need to be exported (clearly visible in Table~\ref{table:wrappers} below). These libraries also tended to use generic names (for example, PETSc represents matrices and vectors with the \texttt{Mat} and \texttt{Vec} types along with a large number of name prefixes; METIS uses \texttt{idx\_t}; a sub-library of Trilinos uses \texttt{MAX\_LINE\_LENGTH}) that, because they are not namespaced, cannot be found using \texttt{grep} in the same way as one can, for example, find all used Kokkos symbols by \texttt{grep}ping for \texttt{Kokkos::}. As a consequence, one has to repeat the edit-compile cycle until the compiler has pointed out, in its error messages, all names declared by a library that need to be wrapped. These libraries also tended to make more use of anonymous \texttt{enum}s and the preprocessor. Examples for this category of hard-to-wrap library are MPI and PETSc.
\end{itemize}

\begin{table}
\caption{\it For each wrapped external dependency of \dealii{}, the table shows how many exportable symbols were exported via a \texttt{using} declaration (column 2); how many non-exportable symbols (either with internal linkage, or preprocessor variables) were wrapped using the methods outlined in the text (column 3); and how many preprocessor macros had to be duplicated (column 4). Note that all of the libraries mentioned define many more names than just the ones listed here; these are only the ones used in \dealii{} and that consequently needed to be wrapped. (References for libraries not already referenced in the introduction: ADOL-C \cite{griewank1996adolc}, CGAL \cite{cgal-user-ref}, HDF5 \cite{hdf5-web-page}, Kokkos \cite{trott2022}, METIS \cite{karypis1998metis}, MUMPS \cite{amestoy2019mumps}, muParser \cite{muparser-web-page}, OpenCascade \cite{opencascade-web-page}, p4est \cite{burstedde2011p4est}, PSBLAS \cite{Filippone2000}, SLEPc \cite{hernandez2005slepc}, SUNDIALS \cite{hindmarsh2005sundials}, Taskflow \cite{huang2021taskflow}, TBB \cite{reinders2007tbb}, UMFPACK \cite{davis2004umfpack}, VTK \cite{vtkBook},
ZLIB \cite{zlib-web-page}.)
}
\label{table:wrappers}
\begin{tabular}{|l|r|r|r|}
     \hline
     \textbf{Dependendency} & \textbf{\# of exported names} & \textbf{\# of wrapped names} & \textbf{\# of duplicated macros} \\ \hline
     C++         & 554 & 7 & -- \\ \hline 
     ADOL-C      & 18 & 1 & -- \\ 
     Boost       & 156 & -- & 1 \\ 
     CGAL        & 71 & -- & -- \\ 
     HDF5        & 47 & 18 & 1 \\ 
     Kokkos      & 56 & 3 & -- \\ 
     METIS       & 5 & 1 & -- \\ 
     MPI         & 88 & 52 & -- \\ 
     MUMPS       & 1 & -- & -- \\ 
     muParser    & 4 & -- & -- \\ 
     OpenCascade & 86 & 2 & -- \\ 
     p4est       & 141 & 3 & 2 \\ 
     PETSc       & 306 & 74 & 16 \\ 
     PSBLAS      & 6 & -- & -- \\ 
     SLEPc       & 38 & 19 & -- \\ 
     SUNDIALS    & 99 & 14 & -- \\ 
     Taskflow    & 7 & -- & -- \\ 
     TBB         & 14 & -- & -- \\ 
     Trilinos    & 184 & 30 & -- \\
     UMFPACK     & 9 & 6 & -- \\
     VTK         & 11 & 0 & -- \\
     ZLIB        & 6 & 5 & -- \\
     \hline
     \textbf{Total} & \textbf{1905} & \textbf{235} & \textbf{20} \\
     \hline
\end{tabular}
\end{table}

Wrapping external declarations in this way is, admittedly, not a great strategy. It was tedious, required re-compiling all library files many times over until all used external classes and functions had been successfully identified and wrapped, and generally took many hours to get right. Table~\ref{table:wrappers} summarizes for each wrapped external dependency some statistics about how many symbols needed to be wrapped in one of the ways mentioned above. The numbers shown there make clear that this was no small effort. At the same time, it is the only solution currently available with the compilers we have. I will comment more on my experience with this step in Section~\ref{sec:experiences}. Beyond these complaints, however, the approach is successful: Because reading already compiled module partition is substantially faster than reading and parsing a long list of header files, using the example of the \texttt{source/base/utilities.cc} file mentioned at the end of Section~\ref{sec:practical-aspects-external-headers-problem}, after wrapping external libraries, compile time is reduced from 3.9 CPU seconds and 426 MB (using the traditional header system) to 2.0 CPU seconds and 346 MB (using modules).

\subsection{Is there an alternative?}
\label{sec:steps-alternative}

The way to convert existing headers and source files so that we can build a C++20 module is clearly the right way to go: At some point in the future, \textit{every} C++ project will be built as a module, and the script-converted files can at that point become \textit{the} source files of the project. Furthermore, once external dependencies start to provide modules of their own, the effort in writing and maintaining the wrapper files discussed in Section~\ref{sec:practical-aspects-external-headers} will also go away. At the same time, the work described above is clearly substantial, and one may ask whether at least for a transition period, one could get away with less work, perhaps forgoing some of the benefits.

Indeed, one can also realize that a project can ``wrap itself'': \dealii{} could simply add a single new file \texttt{dealii.ccm} that looks just like the wrappers of the previous section except that it exports the names of \dealii{}, instead of an external dependency. One could then continue to build the library as has been done for the past decades, and the only addition is the one wrapper file that is compiled into a module and that exports \dealii{}'s names for use in downstream consumers.

This could work: The wrapper would have to \texttt{\#include} every single \dealii{} header file, and list every class, function, and variable name that is part of the library's public interface. The size of this translation unit would likely be substantial, and it would be a maintenance challenge to keep the list of exported names up to date in view of continuing development in the rest of the library. One also does not gain from the compile time improvements outlined at the end of Section~\ref{sec:practical-aspects-external-headers-solution}, and discussed in more detail in Section~\ref{sec:experiences-technical-side-library}. Finally, the scheme does not provide a long-term transition path towards the replacement of header files by module imports. As a consequence, I did not pursue this path, but acknowledge that for other projects -- and for projects that are written in C, Fortran, or other languages, and that want to provide C++ module-based bindings in the same way they currently do with header files -- this alternative path to creating a module is a viable strategy.

\subsection{Connecting it all in CMake}
\label{sec:steps-CMake}

\subsubsection{Building a module for a library.}
\label{sec:steps-CMake-building}
The fact that modules are built from module partitions of different kinds, and the fact that partitions can depend on each other in complex ways, makes it impractical to write a build system for modules by hand -- say, using the traditional \texttt{Makefile} approach. Indeed, tracking dependencies between the many source and header files in traditional C++ is difficult enough to express in hand-written \texttt{Makefile}s; doing so for modules where the dependencies are not expressed in terms of file names but module partition names that need to be mapped to file names is just not manageable for anything other than toy projects. On the other hand, tracking dependencies and determining how and when each file should be compiled can of course be automated with build systems that are able to query compiler output in the same way as the list of included header files can be obtained by invoking the compiler with the \texttt{-M} flag (in the case of GCC and Clang, to give an example). This sort of automation has recently become available; specifically, at the time of writing, there is one cross-platform solution to build C++20 modules: Using CMake as the configuration management system, and Ninja as the build system. Both have gained the necessary support in versions that were release in 2024 (specifically, CMake 3.26 and Ninja 1.11) and that are now available in recent Linux distributions.

Conceptually, CMake makes this step quite simple: Just like for building a number of source files into a (shared or static) library by describing the source files that form the library, one can describe a module by declaring a library, listing its source files, and marking all source files (interface and implementation module units) as a \texttt{FILE\_SET} of type \texttt{CXX\_MODULES}. In only slightly shortened form, this then looks as follows:%
\footnote{There is, of course, a substantial amount of set-up code in addition to the code snippet shown: One needs to say which compiler flags to use, where to find headers and libraries of external dependencies, and much more. But this is the same for the module-based as for the header-based compilation, and thus omitted.}
\begin{codebox}[]{CMakeLists.txt}
    \begin{lstlisting}[language=Gnuplot]
add_library(dealii_module_debug SHARED)
target_sources(dealii_module_debug
               PUBLIC
                 FILE_SET dealii_module_debug_public_sources
                   TYPE CXX_MODULES
                   FILES
                     ${_interface_module_partition_units}       # All converted .h files
                     ${_wrapper_module_units}                   # Wrappers for external libraries
                     ${_primary_module_interface_unit})
target_sources(dealii_module_debug
               PRIVATE
                 FILE_SET dealii_module_debug_private_sources
                   TYPE CXX_MODULES
                   FILES
                     ${_implementation_module_partition_units}) # All converted .cc files
    \end{lstlisting}
\end{codebox}
For reasons that will become clear in Section~\ref{sec:steps-CMake-exporting}, we mark the interface units as \texttt{PUBLIC} sources, and the implementation units as \texttt{PRIVATE} in the two statements above.
For the roughly 1,000 converted header and source files, CMake then constructs targets for the build system (Ninja) that (i)~scan each file for the module partition it exports; (ii)~scan each file for the module partitions it imports, if any; (iii)~build a dependency tree for the order in which module partitions have to be compiled; (iv)~compile each file; (v)~combine the result into a shared library; (vi)~create a file that describes the module's exported interface. A key point here is that the dependency tree can only be created once files have been scanned -- in other words, the build system must be able to dynamically react to the injection of compile rules; it is this ability that has taken until 2024 to develop in Ninja. In total, Ninja has to work through some 3,200 build targets to build the debug-mode shared library, and as many again to build the release-mode one. From a programming perspective, however, the effort to describe building a module is not fundamentally different from that to describe any other shared library.

\subsubsection{Exporting a module.}
\label{sec:steps-CMake-exporting}
At the end of the day, we build software libraries because we want them to be used in applications. To this end, it is necessary to ``install'' the library and everything else applications need to compile and link against the project. For header-based projects, this is straightforward in CMake: Any library declared via \texttt{add\_library(...)} is automatically marked for installation in \texttt{\$CMAKE\_INSTALL\_PREFIX/lib}; one then also marks header files for installation in \texttt{\$CMAKE\_INSTALL\_PREFIX/include}. Importantly, because installed header files are just plain text files, every standards-compliant compiler can understand them, regardless of whether it is the same compiler that compiled the library or not; similarly, because the Application Binary Interface (ABI) of the platform defines the contents and structure of libraries, one can link against installed libraries with few restrictions.

This is not so with modules. To understand why, consider that when compiling an interface partition unit (see Section~\ref{sec:module-outline-interface-unit}) the compiler must produce both an object file that contains compiled code, and a machine-readable file that contains the declarations exported by the partition. The latter is commonly referred to as the ``Built Module Interface (BMI)'' of the partition; the BMIs of individual partitions are later merged into the BMI of the module as a whole upon compiling the primary module interface unit described in Section~\ref{sec:module-outline-primary-unit}. Importantly, the format of BMI files is not standardized and typically changes with every version of the compiler.

How, then, can downstream projects import a module-based library? In order to write \texttt{import dealii;}, it is necessary for the compiler to read the BMI of the module called \texttt{dealii}. In the following, let me discuss the two ways the community currently envisions this to happen; however, I will caution that neither of these is currently well supported by CMake, and there does not appear to be consensus whether either of these is ultimately how this will be done once modules have become a settled matter. CMake may also evolve in this area, and the code snippets below should be understood as conceptual examples rather than actually working code. The discussion below is informed by a number of blog posts, and reports produced as part of the SG15 ``Tooling'' Interest Group of the WG21 C++ Standardization committee; these reports are listed at \url{https://a4z.gitlab.io/blog/2024/11/16/WG21-SG15.html}. Specifically, I have drawn on \cite{ScottBlog,p2409r0,p2473r1} as resources.

\paragraph*{Installing the Built Module Interface (BMI)} The first conceivable approach for making a software library's module available to downstream packages that want to compile against it, is to install the BMI alongside the compiled (shared or static) library and whatever header files are necessary (such as the \texttt{config.h} file or derivatives discussed in Section~\ref{sec:steps-config.h}). The compiler compiling a package that uses this module can then read in this BMI upon seeing \texttt{import dealii;} in the same way as the linker will link against \texttt{libdealii.so}.

The following CMake code snippet (which must be merged with the one in the following section) installs the BMI:
\begin{codebox}[]{CMakeLists.txt}
    \begin{lstlisting}[language=Gnuplot]
install(TARGETS dealii_module_debug
        EXPORT  dealii_module_debug_MIU
        CXX_MODULES_BMI
          DESTINATION "${CMake_INSTALL_LIBDIR}/CMake/dealii_module_debug/bmi"
       )        
    \end{lstlisting}
\end{codebox}

While this seems like the obvious approach at first, it does not work because BMIs do not follow an established file format standard -- in fact, a common approach towards implementing module support in compilers is to serialize the in-memory representation of the set of declarations, which is clearly a compiler-version-dependent format. As a consequence, one needs to use the same compiler and compiler version to consume a BMI as was used to create it in the first place. In some contexts, this is feasible. Often, it is not: For example, a periodic background process updated CLang 20.1.4 to 20.1.5 on my development machine, leading to errors that the previously built BMI is no longer usable. In practice, these issues mean that this approach is likely not usable if one expects a library installed in a system-wide location to be used as the basis of ongoing development in user space, as is the case for many software projects created as part of mathematical research.

\paragraph*{Installing all interface units} The alternative is to install all interface units of the project in the same way as projects currently install header files. A downstream consumer would then first have to compile these interface units into \textit{its own} copy of the BMI, using the compiler used for the consumer, as part of the consuming project's build system.

In simplified form, installing the module interface units (MIUs) via CMake requires the following lines of code which reference the \texttt{dealii\_module\_debug\_public\_sources} set of files previously declared in the code snippet shown in Section~\ref{sec:steps-CMake-building}:
\begin{codebox}[]{CMakeLists.txt}
    \begin{lstlisting}[language=Gnuplot]
install(TARGETS dealii_module_debug
        EXPORT  dealii_module_debug_MIU
        FILE_SET dealii_module_debug_public_sources
          DESTINATION "${CMake_INSTALL_LIBDIR}/CMake/dealii_module_debug/miu"
       )        
    \end{lstlisting}
\end{codebox}
Note how this only installs that set of source files -- the interface units -- that we had previously marked as \texttt{PUBLIC}, but not the implementation partitions that were marked as \texttt{PRIVATE}. This is because only the interface units form the \textit{interface} of the library, and consequently only those are necessary to re-generate the BMI of the library.

Recompiling all interface units in a downstream application to re-generate the BMI is a lot of work. At the same time, it is worth remembering that in the current header-based approach, the consuming project's compiler has to read, parse, and understand \dealii{}'s header files \textit{every time} one compiles a file. In contrast, the consuming project's build system would have to only create the \dealii{} BMI once; every re-compilation would then re-use it when seeing \texttt{import dealii;}. In practice, building the \dealii{} BMI takes about 2.5 minutes of CPU time (to be compared to the roughly 1--2 hours to compile the entire library -- see the ``All major dependencies'' column in Table~\ref{table:compile-time} below); compiling with \texttt{-j6} reduces this to about 30 seconds of wall time. This is unpleasant when compiling an application the very first time; it would certainly be an obstacle if applications had to repeat this for all of their dependencies. For example, the ASPECT code mentioned elsewhere requires 70--80 minutes of CPU time to compile; adding 2.5 minutes to re-compile \dealii{}'s BMI along with hypothetical 11m30s and 12m10s to re-compile the BMIs of its two largest external dependencies Trilinos and BOOST (obtained by simply scaling by the relative sizes of these libraries compared to \dealii{}) would be unacceptable and negate the benefits of modules. I will come back to actual numbers for this approach in Section~\ref{sec:experiences-technical-side-applications} where I actually use this approach to build ASPECT with modules.

In view of this, the ``alternative'' scheme outlined in Section~\ref{sec:steps-alternative} in which a library ``wraps itself'' might appear more attractive: In this scheme, there is only a single interface unit that would have to \texttt{\#include} every single header file of the project, and then re-export all symbols into a module. Whether that turns out to be faster is not clear; surely, including and parsing 600,000 lines of header files is not going to be cheap either. Moreover, the downsides of the approach mentioned at the end of Section~\ref{sec:steps-alternative} of course remain true.

\paragraph*{Summary} At this point, there does not appear to be a consensus on how exactly software packages should install module-based libraries in ways that would allow downstream packages to use them. \cite{p2473r1} and \cite{p2581r2} suggest to install \textit{both} the BMI and the interface partition units, and letting the build system and/or compiler figure out whether the BMI is sufficient or whether the BMI needs to be regenerated. \cite{p2409r0} outlines requirements for the re-use of built modules in downstream projects, but it is not clear whether these are implemented in current compilers and configuration/build systems. In particular, there are no standards such as POSIX that describe where interface units and BMIs are to be installed and found, and what format they have to have. Moreover, while CMake knows how to install both the BMI and the interface partition units, it currently does not actually know how to \textit{use} them in downstream projects: CMake does not have any logic at this point that would read information about installed module artifacts and communicate information about them to a compiler in projects that want to use an upstream project's module. One would therefore have to write the necessary CMake logic oneself that includes, for example, the installed interface units into the set of source files to compile. I will use this approach in Section~\ref{sec:experiences-technical-side-applications}. On the other hand, I build the programs in Section~\ref{sec:experiences-technical-side-steps} from within \dealii{}'s build system, using the BMI compiled as part of compiling \dealii{} itself.

In summary, for now, there is no automatic way to use an installed module. At the same time, one can likely expect that solutions to these problems will emerge over the coming years.

\section{Does it actually work? Practical experience with the resulting conversion}
\label{sec:experiences}

Having described how one can convert large software projects to use C++20 modules, let me turn to a description of how I thought that actually worked. I will divide this section into three parts: Comments about the overall feasibility of converting large-scale mathematical software to a module-based system (Section~\ref{sec:experiences-feasibility}), the technical side that describes technical metrics of success (Section~\ref{sec:experiences-technical-side}), and the human side that covers topics such as how much work it was (Section~\ref{sec:experiences-human-side}).

\subsection{Feasibility of the conversion to C++20 modules}
\label{sec:experiences-feasibility}

My first goal for this study was to assess whether the conversion of a large mathematical software package to a module-based system is possible to begin with. The answer to this is clearly ``yes''. It took a substantial amount of work, see Section~\ref{sec:experiences-human-side}, but it is clear that there is a path for existing, widely-used software packages from their header-based to a module-based interface that downstream applications can use.

\subsection{Technical conclusions from the conversion to C++20 modules}
\label{sec:experiences-technical-side}

In order to compile a project of the size of \dealii{} successfully, one needs three essential components: A reliable configuration management system, a reliable build system, and a reliable compiler. For the initial development, I used CMake 3.28, Ninja 1.11.1, and Clang 20.1 (earlier versions of Clang contained too many bugs to work around).%
\footnote{Following the initial round of review of this paper, I re-did all experiments discussed below and for those upgraded to a pre-release version of Clang 23 (git hash 4e3a074) and CMake 4.3. Compile times and library sizes are therefore reported for Clang 23.}
As mentioned in the introduction, support for C++ modules throughout these tools is relatively recent, and so it was perhaps a surprise that this large-scale experiment did not uncover any bugs in CMake and Ninja. Clang 20 contained a small number of bugs, but one can work around those; one also imagines that these will be fixed in the near future as modules are more widely used. Given that the approach of Section~\ref{sec:steps} is \textit{feasible}, the following subsections then provide metrics to assess whether it is \textit{useful} or \textit{successful}.

\subsubsection{Metric 1: Library compile times}
\label{sec:experiences-technical-side-library}

Using the transformation of header and source files described in Sections~\ref{sec:steps-header-files} and \ref{sec:steps-source-files}, along with the wrapping of external dependencies (Section~\ref{sec:practical-aspects-external-headers}) and the other steps discussed in Section~\ref{sec:steps}, CMake generates a list of 5898 targets for a build with most of the dependencies listed in Table~\ref{table:wrappers}. Ninja then builds these in parallel, using a dependency graph between targets that is automatically generated from both the source files and compiler output obtained from some of the earlier targets. In practice, the majority of the 5898 targets are scanning files for what they export and import, collating this information, and determining dependencies; all of these targets are quickly dispatched. Compiling the roughly 600 module interface units -- which typically do not involve code generation, but only export declarations -- is also relatively quick. The bulk of the build time is therefore spent where it is also for traditional builds: in compiling source files.

One of the measures of defining whether C++20 modules are a success, then, is whether compilation is indeed faster than in the traditional system. Table~\ref{table:compile-time} answers this in the affirmative: On average, compiling the module-based project takes 30--40\% less CPU time compared to the time it takes for the header-based project. The shared libraries that result are about the same size, as has to be expected. Many mathematical software packages are quite large -- Trilinos, with sufficiently many sub-packages enabled, can for example take several hours of CPU time to compile -- and so a 30\% reduction in compile time is a win for our community. The machine used for the experiments shown in the table does not have the memory or file system bandwidth to make full use of the very large core count (when using \texttt{ninja -j128}), but still achieves very nice parallel build speed-ups. Module builds scale slightly worse to large core counts, perhaps due to the need to compile the interface units with their dense dependency graph, but not in a way that would present an obstacle; at lower core counts -- when building with \texttt{-j16} -- wall time for module builds is 25\% faster than for header-based builds.

\begin{table}
\caption{\it A comparison of build times and library sizes between the ``traditional'' and the module builds described herein. \dealii{} builds both debug and release mode libraries. The table shows results for both a minimal set of mandatory dependencies (specifically, only Kokkos and Boost) and when also connecting to nearly all of the optional dependencies shown in Table~\ref{table:wrappers} (specifically, to ARPACK, Boost, GMSH, GSL, Kokkos, LAPACK, MPI, muParser, OpenCascade, p4est, PETSc, SLEPc, SUNDIALS, Taskflow, TBB, Trilinos, UMFPACK, and ZLIB). Builds were all performed on a workstation with two AMD EPYC 7773X 64-core processors and 1 TB of memory. All results shown were obtained by timing \texttt{ninja -j128} and using the best of three runs.}
\label{table:compile-time}

\begin{tabular}{|l|rr|rr|}
  \hline
  & \multicolumn{2}{c}{\textbf{Minimal dependencies}} & \multicolumn{2}{c|}{\textbf{All major dependencies}} \\
  & \multicolumn{1}{c}{Traditional build} & \multicolumn{1}{c}{Module build} & 
    \multicolumn{1}{c}{Traditional build} & Module build  \\ \hline
  & \multicolumn{4}{c|}{Debug build} \\ \hline
  Wall time   & 2m33s &  2m52s & 2m58s & 3m12s \\
  CPU time    & 87m25s & 58m02s & 114m32s & 77m06s \\
  System time &  3m54s &  2m54s &  4m55s &  3m46s \\
  Library size&  571MB &  561MB &  782MB &  769MB \\
  \hline
  & \multicolumn{4}{c|}{Release build} \\ \hline
  Wall time   &  1m38s &  1m53s & 2m06s & 2m22s \\
  CPU time    & 69m12s & 41m31s & 94m45s & 58m36s \\
  System time &  3m14s &  2m12s &  4m10s &  2m52s \\
  Library size&  176MB &  188MB &  272MB &  289MB \\
  \hline
\end{tabular}
\end{table}

\subsubsection{Metric 2: Compile times for small- to medium-sized projects}
\label{sec:experiences-technical-side-steps}
Intuitively, one would expect the largest compile-time gains from the use of modules for files that have relatively little code on their own but \texttt{\#include} a large number of header files -- in other words, high-level code building on the abstractions provided by libraries. Most codes building on \dealii{} actually fall into this category: They have perhaps a few thousand lines of code, but \texttt{\#include} substantial portions of \dealii{} and perhaps (directly or indirectly) also significant parts of libraries such as Trilinos or Boost. 

\begin{figure}
    \centering
    \includegraphics[width=0.9\linewidth]{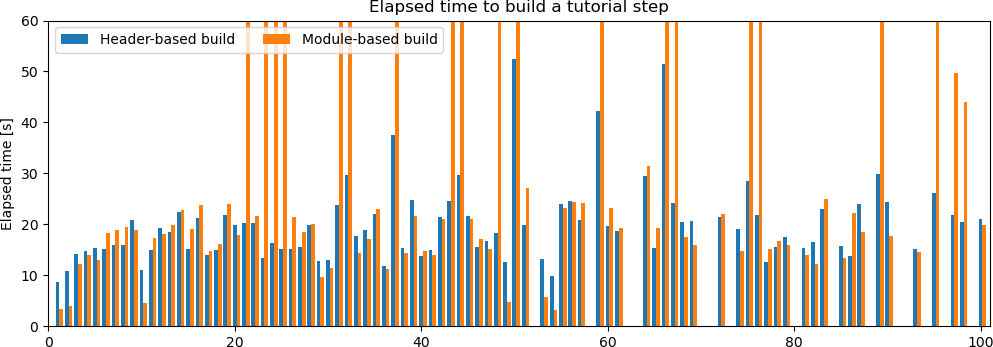}
    \caption{\it Build times for those 86 of the \dealii{} tutorial steps that could be compiled and linked with the dependencies used in this manuscript, using both the traditional header-based approach and the module-based one described herein. Build times for those 18 module-based builds that exceed the shown vertical range are between 73 and 184 seconds. (Numbers on the $x$-axis identify the tutorial program in question: between step-1 and step-100.)}
    \label{fig:compile-time-steps}
\end{figure}

To test this hypothesis, I have used the set of currently 90 tutorial programs that are part of \dealii{} -- referred to as step-1, step-2, etc. -- that have between 280 and 4048 lines of code, and between 30 and 796 lines that contain a semicolon. Only 86 of these can be tested; the rest require an external dependency that was not available on my system.
Figure \ref{fig:compile-time-steps} shows compile times for header-based compilation of the tutorial steps as well as for slightly modified versions that are compiled against the \texttt{dealii} module.

Of the 86 programs, 37 build faster in a module-based build than in a header-based build, with an average improvement of 21\%. On the other hand, the remaining 49 are slower to build in the module-based approach, and on average these take 223\% as much time to compile -- most have comparable times, but 18 are substantially slower, including one that took 7.3 times as long to build this way. There does not seem to be any obvious commonality between the programs that are slower, making the cause for the slowdown unclear for now.

\subsubsection{Metric 3: Compile times for large projects}
\label{sec:experiences-technical-side-applications}
At the other end of the spectrum reside large applications that build on libraries such as \dealii{}. In order to test the effect of modularization on compile time for these applications, I converted the Advanced Solver for Planetary Evolution, Convection, and Tectonics (ASPECT) previously already mentioned \cite{KHB12,HDGB17}. ASPECT has around 200,000 lines of code and consists of 391 header and 394 source files that I converted to C++20 module partition units in a way very similar to the one outlined in Section~\ref{sec:steps}.

ASPECT's code base is very structured: Nearly everything is a plugin (i.e., a class derived from some 25 abstract base classes) and despite its size, the vast majority of files have very similar layouts. As a consequence, the conversion did not actually require much work beyond writing a small number of scripts that apply the same action over and over to many files. In the end, it took only around 25 pull requests to the ASPECT code base, and 30 or so additions of wrapped symbols beyond those listed in Table~\ref{table:wrappers}. All necessary changes to modularize this code and let it import the \dealii{} module instead of including header files were made over the course of six hours of work in a single day.

\begin{table}
\caption{\it A comparison of build times for the Advanced Solver for Planetary Evolution, Convection, and Tectonics (ASPECT) between ``traditional'' and module-based builds. The module-builds also require re-constituting the built module interface (BMI) of the \dealii{} library from \dealii{}'s interface module units (see Section~\ref{sec:steps-CMake-exporting}). This step only has to be done once, and the time is included in the ``Module build'' columns. On the other hand, when changing ASPECT files, one only has to re-compile the latter, but not the \dealii{} BMI; the corresponding build times for this step are listed in the ``[Module] re-build'' column.
Builds were performed on the same machine as mentioned in Table~\ref{table:compile-time}, again using \texttt{ninja -j128}.}
\label{table:compile-time-aspect}
\begin{tabular}{|l|rrr|rrr|}
  \hline
  & \multicolumn{3}{c|}{\textbf{Debug build}} & \multicolumn{3}{c|}{\textbf{Release build}} \\
  & \multicolumn{1}{c}{Traditional build} 
  & \multicolumn{1}{c}{Module build} 
  & \multicolumn{1}{c|}{re-build} 
  & \multicolumn{1}{c}{Traditional build} 
  & \multicolumn{1}{c}{Module build}
  & \multicolumn{1}{c|}{re-build} 
  \\ \hline
  Wall time      &  2m00s  & 4m29s & 4m04s &  1m28s &  3m26s &  3m12s \\
  CPU time       & 81m13s  &35m20s &33m00s & 72m40s & 28m14s & 25m53s \\
  System time    &  5m28s  & 4m35s & 3m42s &  5m11s &  4m11s &  3m11s \\
  \hline
\end{tabular}
\end{table}

Table~\ref{table:compile-time-aspect} shows how long it took to compile ASPECT, in both debug and release configurations. Because this build is outside the build directory of \dealii{}, I included \dealii{}'s installed module interface units into the list of source files of ASPECT in order to let the compiler re-constitute \dealii{}'s BMI (see Section~\ref{sec:steps-CMake-exporting}). This step clearly takes time (some 2m20s of CPU time), and consequently the table also lists how long it takes to compile \textit{only} ASPECT, once the BMI has already been re-constituted. The numbers shown in the table are encouraging: Even including the BMI re-constitution, module-based builds take only around 40\% as much CPU time as header-based builds. This is unconditionally a good thing. The only downside is that with such highly parallel builds (using \texttt{-j128}), wall time is larger. With \texttt{-j16}, the wall time for a module build is 10\% \textit{faster} than for a traditional build. The underlying reasons for this observation are not entirely clear but probably include that compiling all interface units (i.e., header files) with their mutual dependencies limits the amount of parallelism compared to only having to compile source files that are all independent of each other.

\subsubsection{Metric 4: Code structure and maintainability}
\label{sec:experiences-technical-side-code}
A separate technical measure is whether the code base as a whole is in better shape. This is a more mixed bag. On the one hand, the effort of converting the code base has undoubtedly led to cleaning up many places that used idiosyncratic constructs (see also Section~\ref{sec:experiences-human-side} below). For example, module builds do not allow having \texttt{static} functions in interface partitions; yet, in both our code base and that of several of our dependencies, one can find many \texttt{static} functions in header files. Such functions are duplicated in every \texttt{.cc} file that includes this header file, and so marking functions in header files as \texttt{static} was \textit{never} a good idea. Cleaning these places up is therefore a positive.  Moreover, none of the changes I had to make to the \dealii{} code basis led to any incompatibilities with previous versions.

On the other hand, one of the major efforts of the conversion was to write wrappers for all of the external dependencies listed in Table~\ref{table:wrappers}. This added around 5,300 lines of code to the library, an effort I found not only exceedingly tedious, but also rather undesirable because it replicates the interface of external libraries. Since \dealii{} -- like many other mathematical software packages -- needs to be able to interact with whatever version of an external dependency is found on a user's system, this implies that the wrappers also track the various interface changes of other packages; for example, the export list of the Kokkos wrappers looks as follows:
\begin{codebox}[]{kokkos.ccm}
\begin{lstlisting}
export
{
  namespace Kokkos
  {
    using ::Kokkos::abort;
    using ::Kokkos::abs;
#if DEAL_II_KOKKOS_VERSION_GTE(4, 0, 0)
    using ::Kokkos::ALL;
#endif
    [...]
\end{lstlisting}
\end{codebox}
Many of the other wrappers look similar. The need to account for version changes is something that seems to set mathematical software apart from more widely used packages such as Boost: Mathematical software packages are generally less widely used than, say, Boost and with the smaller number of users comes a greater willingness to change interfaces on a regular basis. (As mentioned in the introduction, having to deal with external dependencies with no control over how they change also separates mathematical software from many commercial code bases.) Keeping wrappers up to date will consequently no doubt be an ongoing effort. On the other hand, one can hope that perhaps the work invested in writing these wrappers will eventually help the authors of the external packages to export their functionality in the form of C++20 modules as well -- the code described herein is available under an open source license for others to use.

\subsubsection{Metric 5: Effect on downstream projects}
The data of the previous subsections present \textit{some} evidence that building software as C++20 modules is useful. The part that is currently unsolved is how downstream projects can benefit from this: As discussed at length in Section~\ref{sec:steps-CMake-exporting}, there is currently no easy or uniform way to export a project's modules to downstream users, and in particular no easy way to \textit{use} any installed module artifacts. While it is not difficult to manually tell CMake to include installed interface partition units into the set of source files to build (as I have done in my experiment with ASPECT, in Section~\ref{sec:experiences-technical-side-applications}), an only slightly simplified answer to the question of this section -- ``Does it actually work?'' -- is ``No, not for downstream projects, and at least not automatically''. 

More specifically, to make it work one needs to coordinate knowledge about file sets and installation directories between the upstream producer and the downstream consumer of a module if one wants to add the module interface units to the source files of the downstream user. There is no support for re-using a built module interface (BMI), and so there is no way around the price to pay for re-compiling it.
As a consequence, at least for now, modules are most useful for the terminal applications of a dependency tree between software libraries and applications.

\subsection{The human side of converting to C++20 modules}
\label{sec:experiences-human-side}

The way modules are described in C++20 -- as outlined in Section~\ref{sec:module-outline} -- seems rather clunky. Compared to the ease with which one creates a module or package in Python by simply placing source files into a directory, for example, doing so in C++20 requires quite a lot of writing. At the same time, I found that what C++20 provides is actually pretty well aligned with what is necessary to convert large projects like the one mentioned herein. After an initial phase of learning the terminology of modules (say, the idea of module partitions, what interface and implementation units are, what and how declarations are exported and imported), I did not run into areas where I could not figure out how existing code or code structure should be mapped onto the ideas of the module system. In other words, I did not run into \textit{conceptual} problems that I could not resolve.

In practice, of course, the actual conversion was nevertheless \textit{a lot} of work: In the range of 1.5 months of full time work. As mentioned in Section~\ref{sec:principles}, the only reasonable way to convert large software projects is to write scripts that convert all files of the same kind automatically. These scripts are not extensive -- collectively perhaps 150 lines of Python, plus comments and long lists of header files corresponding to external packages, all written in what was probably not much more than two or three hours. Writing the code to connect everything via CMake took perhaps another 4-6 hours. In theory, the conversion to using modules based on these scripts is then relatively straightforward if all header and source files conform to the expectations of the scripts. In actuality, of course, this is not the case, and the vast majority of time was spent on (i) wrapping external projects, and (ii) dealing with ``special'' cases of files that trip up scripts or compilers. One can deal with these special cases in three different ways:
\begin{itemize}
    \item In a small number of cases, compilation problems can be dealt with by generalizing scripts. For example, nearly every symbol that \dealii{} declares is in the \texttt{dealii} namespace, and these cases are efficiently dealt with by keying the conversion off the \texttt{DEAL\_II\_NAMESPACE\_OPEN} and \texttt{DEAL\_II\_NAMESPACE\_CLOSE} markers discussed in Section~\ref{sec:steps-header-files}. But there are also some 20 files that place declarations into the global namespace, into namespace \texttt{std}, or into the namespaces of other packages \dealii{} interfaces with. Because we need to export these declarations into the module, but because these files also have to continue to compile in non-module mode, I dealt with them by introducing ways in the conversion scripts to mark exported sections that are not tied to pairs of \texttt{DEAL\_II\_NAMESPACE\_OPEN}/\texttt{DEAL\_II\_NAMESPACE\_CLOSE} markers.
    
    \item In a moderate number of cases, the existing header or source files were simply wrong, or perhaps not very robust, and in those cases it is to the general benefit of the project to just fix the original sources. As an example, the \dealii{} source files have 6478 \texttt{\#include} statements that use the form \texttt{<deal.II/directory/file.h>}, but there were also 19 that use the syntax \texttt{"deal.II/directory/file.h"} that predictably led to compiler errors on the converted files because the script did not convert the latter syntax to \texttt{import} statements. In other cases, header files exported constants as variables marked with the `\texttt{static}' keyword, which is allowed (but not a good solution) for traditional header files, but not allowed for modules. In yet other cases, header files cyclically included each other -- something that is not wrong if one uses appropriate include header guards, but that is a brittle design. Modules, however, do not allow cyclic dependencies between modules.
    There were many other cases of this sort in our code basis, as there likely are in every sufficiently large project.
    
    In the mentioned situations as well as related ones, the solution was to simply fix the existing code through a long sequence of pull requests against the main branch of the software.

    \item Finally, a large number of cases simply required manual and often repetitive labor. Examples include the addition of more than 500 \texttt{\#include} statements to around 200 \dealii{} files (see the discussion of transitive inclusion in Section~\ref{sec:steps-header-files}), and in particular writing wrappers for all external dependencies (Section~\ref{sec:practical-aspects-external-headers}). Writing the wrappers for external dependencies alone likely took the equivalent of a week of work. This is simply not glorious work, neither to write nor to review.
\end{itemize}
In the end, however, 1.5 months of work is not an overwhelming amount of work for a project as large as \dealii{} -- the \texttt{sloccount} program estimates that the development of the code in the \texttt{include/} and \texttt{source/} directories alone took 120 person years of work, with the development of all of \dealii{} substantially beyond that. Moreover, while many aspects of the work required a very solid knowledge of C++, none of the work was difficult (say, rewriting substantial portions of code) or so boring that it simply did not make sense to do. To get a sense of the work that needed to be done, the reader may consult the more than 250 pull requests linked to from the github issue that tracks my work, see \url{https://github.com/dealii/dealii/issues/18071}).

\section{Conclusion and an outlook on how the mathematical software ecosystem might evolve}
\label{sec:conclusions}

Many of the most widely used mathematical software packages are written in C++, a language that since its invention in 1985 has continued to evolve. In order to stay relevant (or simply to attract younger programmers to the project), these packages have to co-evolve as well, and one of the areas in which that may be most difficult is the conversion towards using the module system that was introduced with C++20, given that the transition can not be achieved gradually via small changes here and there. Using C++20 modules promises a more predictable way to export functionality to downstream packages, as well as to lower the cost of compilation. It also introduces a feature that programmers are used to from other programming languages (say, Python) they may have encountered before C++.

This contribution summarizes my attempts at finding out what it takes to achieve such a conversion on a large mathematical software package, the \dealii{} library, with its around 800,000 lines of code not including the test suite. As discussed, one can convert a library of this size in a manageable time frame (around 1.5 person months of work), without having to rewrite large parts of the code or introducing incompatible changes. While the evidence shown above is pretty clear that building a software package as a module provides the claimed benefits in terms of compile time (a reduction by 30-40\% for the library itself, see Section~\ref{sec:experiences-technical-side-library}, and even more for the large application in Section~\ref{sec:experiences-technical-side-applications}) and perhaps better code structure (Section~\ref{sec:experiences-technical-side-code}), the data shown in Section~\ref{sec:experiences-technical-side-steps} also make clear that the effect on compile time of smaller downstream projects is at best unclear. Moreover, the infrastructure for installing and using modules does not currently exist in configuration management systems such as CMake, though one may hope that it will be provided in the near future. From a practical perspective, the only reasonable approach to converting existing projects is to export \textit{everything} that a package currently declares in its header files because it is not feasible to annotate the thousands of declarations whether or not they should be exported as part of the conversion; however, the experience outlined herein may perhaps spur projects to gradually move their code bases towards a style in which those classes and functions that are intended to form the public interface are annotated accordingly.

It is perhaps interesting to speculate how the community of authors of mathematical software packages might best approach the conversion of our ecosystem towards modules in the coming years. As discussed in the introduction, an important feature of mathematical software is the deep and complex web of independent packages building on each other -- as opposed to the ``monorepos'' often used in industry. First, it is clear that packages will have to offer header-based interfaces for a long time to come -- though as discussed herein, it is possible to build both header-based and module-based interfaces from a single code base without introducing incompatibilities. Second, one conceivable approach is a ``top-down'' conversion  of our ecosystem in which applications first start to use modules, wrapping the libraries they depend on as discussed in Section~\ref{sec:practical-aspects-external-headers}; as the highest libraries are converted as well, applications' own wrappers can then simply be replaced by the modules exported by these libraries, and the process continues to lower and lower levels of the dependency stack. This top-down conversion requires writing wrappers -- an unpleasant prospect given how long this took for those libraries covered by Table~\ref{table:wrappers}, although one might be able to build on others' work wrapping the same underlying library. Third, the alternative is a ``bottom-up'' approach in which the bottom-most libraries -- say, at the level of BLAS -- convert first, building up the dependency tree. The conversion of ASPECT detailed in Section~\ref{sec:experiences-technical-side-applications} shows that once a module interface for a dependency library is available, the conversion of a higher-level code can actually be quite easy. Ultimately, the only truly bad outcome in the long run is if some projects are neither converted nor wrapped whereas others are, given that header files and module imports do not coexist well as outlined in Section~\ref{sec:practical-aspects-external-headers-problem}.

Of course, in practice, the many projects that occupy the mathematical software scene are made up of individual people, subject to personal interests and funded by a large number of agencies that have differing goals. As a consequence, each project will individually decide whether and when to convert to the use of C++ modules. Their decisions will be independent, though they will perhaps be influenced in their timing by considerations such as the ones that compare the top-down and bottom-up approaches. In the end, we can hope that our ecosystem will continue to evolve -- rather than stagnate -- and adapt to the continued evolution of C++.

\subsection*{Acknowledgments}

\dealii{} is a community project run by a large group of Principal Developers, and encompassing the contributions of nearly 400 people from around the world over the course of more than 25 years. Reporting on my experiences would not have been interesting had it not been for the effort all of these people put into \dealii{} over the years, making it a project that is relevant to many in the scientific computing community.

I would like to specifically acknowledge the work of my fellow Principal Developers who reviewed the more than 250 pull requests that make up the work reported herein. I appreciate their patience in reviewing, providing feedback, and merging pieces that individually all did not actually provide functionality or make anything about the software better until the final stone fell in place!

I would also like the acknowledge the guidance of Clang developer Chuanqi Xu, who pointed me at the ideas related to wrapping external projects as discussed in Section~\ref{sec:practical-aspects-external-headers}. Finally, the work discussed herein has led to patches to numerous other projects among \dealii{}'s external dependencies, mostly where structures used in their header files do not play well with wrappers. These projects have all been open to pull requests that morph their code base towards what I needed herein, even though the existing code worked just fine for their needs.

\subsection*{Funding}

The author's work was partially supported by the National Science Foundation under awards EAR-1925595 and OAC-2410847.

% --------------------
% --- Bibliography ---
% --------------------

\bibliographystyle{ACM-Reference-Format}
\bibliography{paper.bib}

\end{document}